\documentstyle[aps,prl,twocolumn,floats,graphicx]{revtex}

\begin{document}
\draft \preprint{}
\title{Optical conductivity of a superfluid density wave}
\wideabs{
\author{Joseph Orenstein}
\address{Materials Sciences Division, Lawrence Berkeley National Laboratory and Physics Department, University
of California, Berkeley, California 94720}
\date{\today}
\maketitle
\begin{abstract}
We present a calculation of the low frequency optical conductivity
of a superconductor in the presence of quenched inhomogeneity in
both the superfluid and normal fluid densities.  We find that
inhomogeneity in the superfluid density displaces spectral weight
from the condensate to a frequency range that depends critically
on the spatial correlation of normal and superfluid density
fluctuations.
\end{abstract}
}

Measurements of the optical conductivity, $\sigma(\omega)$, in the
microwave and terahertz regimes have proved to be a valuable probe
of quasiparticle dynamics in cuprate superconductors. Because even
relatively poor samples are in the clean limit where the
scattering rate is smaller than the gap frequency, or
$1/\tau\ll\Delta$, the spectral weight of the dissipation due to
quasiparticle scattering is far greater than that associated with
pair creation. As a consequence the quasiparticle Drude peak can
be clearly resolved, and its width can yield the scattering rate.
Important clues as to the nature of the scattering can be obtained
by comparing the rates measured by microwave conductivity with
those measured by angle-resolved photoemission (ARPES).

Extracting $1/\tau$ from the microwave conductivity of a cuprate
requires $\sigma(\omega)$ to be consistent with the two-fluid
model for a clean, d-wave superconductor \cite{bonn}.  In this
model the total conductivity contains two contributions: a
$\delta$-function at $\omega=0$ due to the condensate and a
Drude-like peak due to quasiparticle scattering. The spectral
weights of these components are the super and normal fluid
densities, respectively. As the temperature $T$ tends to zero the
normal fluid density, $\rho_n$, is expected to decrease linearly
with $T$. The superfluid density, $\rho_s$ is expected to increase
at the same rate, so that the total density remains constant. This
simple behavior permits $\rho_{n,s}$ and $1/\tau$ to be determined
in the superconducting state.

The two-fluid model provides an excellent description of the
microwave properties of $YBa_{2}Cu_{3}O_{7-\delta}$ (YBCO) single
crystals \cite{hosseini}. The scattering rates inferred from the
two-fluid analysis are remarkably small, approaching $\sim$ 10 GHz
as $T\rightarrow0$. On the other hand, analysis of ARPES yields
low temperature scattering rates that are larger by a factor of at
least 100 \cite{valla}.  One possible explanation is that the
single particle self-energy measured by ARPES is very different
from the momentum relaxation rate that determines the
conductivity.  However, an alternative explanation is that the
ARPES data are collected almost exclusively from studies of
$Bi_2Sr_2CaCu_2O_{8+\delta}$ (BSCCO) rather than YBCO.

It would seem that the microwave data on BSCCO would hold the
resolution to this question. However, analysis of such data has
been problematic because the spectra are not consistent with the
two-fluid model described previously. In fact the success of the
two-fluid description is unique to YBCO single crystals - all
other cuprates that have been measured show distinctly different
properties \cite{sflee,jacobs,pronin,waldram}. In these materials
the spectral weight of the Drude peak extrapolates to a large
residual value as $T\rightarrow0$, suggesting that a substantial
fraction of the normal fluid fails to condense.

Recently, the frequency dependence of the conductivity was
reported in optimally doped BSCCO \cite{corson}. The measurements
were performed using coherent time-domain spectroscopy, a
technique which can determine $\sigma(\omega)$ up to frequencies
$\omega/2\pi\sim$1 THz. The width of the Drude peak at low
temperature was determined to be $\sim$300 GHz. Integration of the
real part of the conductivity over the experimental range showed
that the uncondensed portion is at least 30$\%$ of the condensate
spectral weight.  A study of BSCCO for several carrier
concentrations showed that this percentage increases with doping,
reaching about 60$\%$ in an overdoped sample \cite{phd}.

The two-fluid description of $\sigma(\omega, T)$ in such samples
is not internally consistent. Fitting the spectra requires
$\rho_n$ to vary with frequency and actually increase with
decreasing $T$ for $\omega \lesssim $300GHz. On the other hand,
the quasiparticle conductivity could be described by physically
reasonable parameters if Re $\sigma$ has a third component: a peak
centered at $\omega$=0, whose spectral weight grows in proportion
to $\rho_s$ as $T$ decreases.

The proportionality to $\rho_s$ suggests that this anomalous part
of $\sigma$ is related to current fluctuations of the superfluid.
However, in a homogeneous superconductor the spectral weight of
such fluctuations decreases exponentially with $T$, which is
opposite to the growth of spectral weight that was observed.
Recently, Barabash \textit{et al.} \cite{bs} showed that a
component of Re $\sigma$ can increase as $T\rightarrow0$ if
$\rho_s$ is spatially inhomogeneous.  They considered a Josephson
junction network with a quenched variation $\delta J$ about the
mean coupling $\bar{J}$. ($J$ is the superfluid density defined on
a lattice rather than a continuum). They showed that the global
phase stiffness $J$ equals $\bar{J}-\langle \delta
J^2\rangle/\bar{J}$. $J$ is less than $\bar{J}$ because the phase
varies more rapidly in regions where the stiffness is below the
average value.

The result stated above is important to the optical conductivity
because the spectral weight of the condensate $\delta$-function is
proportional to $J$, whereas the total spectral weight of
condensate current fluctuations is proportional to $\bar{J}$. Thus
inhomogeneity of the superfluid density must displace spectral
weight $\sim \langle \delta J^2\rangle/\bar{J}$ from the
$\delta$-function to nonzero frequency. If $\delta J$ varies in
proportion to $\bar{J}$ then the displaced spectral weight will
track that of the condensate, in agreement with terahertz
experiments \cite{corson}.

The existence of strong spatial inhomogeneity in BSCCO has been
demonstrated by scanning tunneling microscopy (STM) measurements
\cite{pan}. The local density of states (LDOS) of BSCCO varies
randomly in space, with spatial fluctuations that have a minimum
wavelength of $\sim$50 $\AA$.  The variations in LDOS suggest
quenched inhomogeneity in local carrier concentration, $x$, and
therefore in the local $\rho_s$.

To determine whether quenched inhomogeneity in $\rho_s$ is
responsible for the anomaly in $\sigma(\omega)$, we need to
consider the spectrum of the spectral weight removed from the
condensate. In a system with no normal component the spectral
weight would be expected to appear at the natural oscillation
frequency of the condensate, which is the Josephson plasma
frequency, $\omega_s$. In optimal cuprates
$\omega_s/2\pi$$\sim$200 THz, whereas the anomalous dissipation is
found below $\sim$1 THz. Inhomogeneity can only explain the data
if the presence of a normal fluid in addition to the superfluid
causes the displaced spectral weight to appear at frequencies much
smaller than $\omega_s$.

In this paper we calculate the change in $\sigma(\omega)$ due to
inhomogeneity in a superconductor in which normal and superfluid
coexist. This is the first calculation in which the spectral
weight of the conductivity is consistent with the theorem of Ref.
\cite{bs}.  We find that $\sigma(\omega)$ is extremely sensitive
to correlations of the normal and superfluid density variations in
the the medium. The spectrum depends crucially on whether the
correlation is positive (such that regions of large $\rho_s$ have
large $\rho_n$) or negative.  If the correlation is positive, the
displaced spectral weight indeed shifts to the plasma frequency.
However, when the fluctuations of normal and superfluid density
are anticorrelated, the spectral weight appears at low
frequencies, in agreement with microwave and terahertz
measurements on cuprate superconductors.

To treat the conductivity in the presence of inhomogeneity we
apply the extended two-fluid phenomenology developed by Pethick
and Smith \cite{pethick}, and Kadin and Goldman \cite{kadin}. This
approach successfully describes quenched inhomogeneity at the
normal-superconductor interface and fluctuating inhomogeneity, as
in the Carlson-Goldman oscillations \cite{cg}. In the extended
two-fluid model the superfluid is accelerated by gradients of the
chemical potential as well as electric fields, that is,

\begin{equation}
\dot{\vec{J_s}}=\rho_s(\vec E-\nabla\mu_s). \label{eq:first}
\end{equation}
The chemical potential has the subscript $s$ because in a
superconductor $\mu$ is the energy per electron required to add a
pair to the condensate.  The corresponding equation for the normal
fluid current requires solving the Boltzmann equation for the
quasiparticle distribution function. However for frequencies less
than $1/\tau$, the distribution function is the equilibrium
distribution shifted by the 'quasiparticle chemical potential' or
$\mu_n$.  If the normal fluid is in local equilibrium with the
condensate $\mu_n=\mu_s$, which differs from "global" equilibrium
where $\mu_n$=0. In the low frequency regime the constitutive
relation for the normal fluid has the simple form:

\begin{equation}
\dot{\vec J_n}=\rho_n(\vec E-\nabla\mu_n)-\vec J_n/\tau.
\label{eq:second}
\end{equation}

A closed system of equations requires continuity relations. The
total charge of the superconductor separates naturally into a
normal component, $Q_n$ that depends on both the coherence factors
and the distribution function,

\begin{equation}
Q_{n}\equiv \sum_{k} q_k f_k \label{eq:fourth}
\end{equation}
where $q_k^2\equiv u_k^2-v_k^2$, and a superfluid component that
depends only on coherence factors,

\begin{equation}
Q_{s}\equiv \sum_{k} 2ev_k^2=2eN_{F}\mu_s. \label{eq:fifth}
\end{equation}
Under conditions for which $\mu_n$ can be defined, the normal
fluid charge is given by,

\begin{equation}
Q_{n}=2N_{F}\lambda(\mu_s-\mu_n),\label{eq:fourth}
\end{equation}
where $N_F$ is the normal state density of states at the Fermi
level. The parameter $\lambda$ relates the normal fluid charge to
the shift of $\mu_n$ away from local equilibrium.

In a superconducting medium the normal and superfluid charge are
not separately conserved. Interconversion of normal and
superconducting charge occurs through two types of processes.  In
the first process, $Q_n$ changes as quasiparticles recombine or
scatter. In the second, the quasiparticle charge changes even if
$f_k$ remains constant. In this process, $Q_n$ varies because the
quasiparticle excitation spectrum, and consequently the effective
charge, adjusts to the local value of $\mu_s$. Continuity
equations that include both types of exchange between the two
fluids are:

\begin{equation}
\dot{Q}_{n,s}+\nabla\cdot \vec
J_{n,s}=(-,+)({Q_n\over\tau_Q}-\lambda\dot{Q}_s),
\label{eq:fourth}
\end{equation}
where $\tau_Q$ is the rate of conversion of normal charge into
superfluid charge due to scattering and recombination processes.
The above system of equations is closed by $\nabla\cdot \vec
E={Q/\epsilon_0}$.

To see how quenched inhomogeneity affects the optical
conductivity, we consider the simplest possible model: a
one-dimensional sinusoidal variation in normal and superfluid
density.  We assume that the densities of the two components vary
as
$\rho_{s,n}(x)=\langle\rho_{s,n}\rangle+Re\{\rho_{sq,nq}e^{iqx}\}$,
where $\langle\rho\rangle$ is the average density and $\rho_q$ is
the amplitude of the inhomogeneity at wavevector $q$. Because the
medium is inhomogeneous, a uniform applied field generates a field
at $q$. Solving the extended two-fluid equations to lowest order
in $\rho_{nq,sq}$, we obtain,

\begin{equation}
{E_q\over E_0}=-{\rho_{sq}+\rho_{nq}F\over\omega_s^2
-\omega^2(1-\lambda)+(\omega_n^2-i\omega\lambda/\tau)F},
\label{eq:fourth}
\end{equation}
where,

\begin{equation}
F\equiv{\omega (\omega+i/\tau_Q^*)-v_s^2q^2/(1-\lambda)\over
(\omega+i/\tau)(\omega+i/\tau_Q^*)-v_n^2q^2/\lambda},
\label{eq:fourth}
\end{equation}
$\omega_{s,n}^2\equiv\rho_{s,n}/\epsilon_0+v_{s,n}^2q^2$,
$v_{s,n}^2\equiv\rho_{s,n}/2N_Fe^2$, and
$\tau_Q^*\equiv\tau_Q(1-\lambda)$.

The uniform current density in response to these fields is given
by $J_0=\sigma_0E_0+\sigma_qE_{-q}$, where $\sigma_0$ is the
uniform two-fluid conductivity, and $\sigma_q$ is the conductivity
that varies with wavevector $q$. In the two fluid model these are
given by,

\begin{equation}
\sigma_{0,q}={i\rho_{s,sq}\over\omega}+{i\rho_{n,nq}\over\omega+i/\tau}.
\label{eq:fourth}
\end{equation}
The effective conductivity of the medium,
$\sigma=\sigma_1+i\sigma_2$, is the ratio of the uniform current
density to the uniform field, so that,
$\sigma=\sigma_0+\sigma_qE_{-q}/E_0$. The second term in this
equation is the change in the optical conductivity due to
inhomogeneity, or $\Delta\sigma$.

The extra term in the conductivity is particularly simple if
$\rho_n=0$, in which case,

\begin{equation}
\Delta\sigma_2=-{\rho_{sq}\over\omega}{\rho_{sq}\over\omega_s^2-\omega^2}.
\label{eq:fourth}
\end{equation}
The inhomogeneity in the superfluid density indeed removes
spectral weight $\rho_{sq}^2/\rho_s$ from the condensate
$\delta$-function, in agreement with the results of Ref.
\cite{bs}. In the absence of a normal fluid component the spectral
weight reappears in a $\delta$-function at the Josephson plasma
frequency.

We next assume that $\rho_n\neq 0$ (even at $T=0$), and describe
how this assumption affects $\Delta\sigma(\omega)$.  We focus on
the behavior of $\Delta\sigma$ when the normal fluid density
fluctuations are either perfectly correlated or anticorrelated
with those of the superfluid density. We take for two-fluid
parameters values that are suggested by the terahertz and
microwave experiments: $(\rho_s/\epsilon_0)^{1/2}$=1000 THz,
$(\rho_n/\epsilon_0)^{1/2}$=800 THz, and $\tau^{-1}=\tau_Q^{-1}$=3
THz. The existence of normal fluid at $T=0$ implies a large
density of states at the Fermi level, $N_0$. While the origin of
$N_0$ is outside the scope of this paper, one may speculate that
it is closely related to the strong inhomogeneity observed by STM.
If we make the reasonable approximation of neglecting BCS
coherence factors for the quasiparticle states introduced by
disorder, then $\lambda=N_0/N_F$.

\begin{figure}[h]
     \includegraphics[width=3.0in]{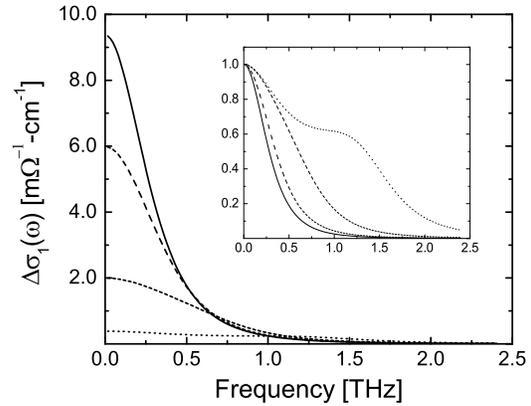}
\caption{$\Delta\sigma_1$ as a function of frequency
($\omega/2\pi$) for anticorrelated variations in $\rho_s$ and
$\rho_n$.  Spectral weight decreases for increasing $v_sq\tau$:
0.25, 0.5, 1.0, and 2.0. The same curves are shown in a normalized
plot in the inset.} \label{fig:First}
\end{figure}

We begin with the case where the density fluctuations are
perfectly anticorrelated, so that $\rho_{sq}=-\rho_{nq}$ and the
total fluid density is uniform throughout the medium. Fig. 1 shows
$\Delta\sigma_1(\omega)$ with $(\rho_{sq}/\epsilon_0)^{1/2}$=100
THz, for several values of $v_sq\tau$. $\Delta\sigma_1(\omega)$ is
positive and centered at $\omega=0$ rather than $\omega_s$. The
spectra depend strongly on $v_sq\tau$. For $v_sq\tau\ll$1,
$\Delta\sigma_1$ has a Drude-like spectrum, whose width is $\sim
1/\tau$. As $v_sq\tau$ increases beyond unity, the spectral weight
drops and a peak near the Carlson-Goldman frequency $v_sq$ appears
in the spectrum.

The key issue is the fraction of the displaced condensate spectral
weight that appears in the low-frequency peak, as opposed to
$\omega\sim\omega_s$. In Fig. 2 we compare the reduction in
condensate weight with the increase in dissipation at low
frequency. The change in condensate spectral weight was evaluated
from the $\lim_{\omega\rightarrow0}(\pi/2)\omega
\Delta\sigma_1(\omega)$. The low frequency spectral weight was
obtained by numerically computing the integral of $\Delta\sigma_1$
with respect to $\omega$ from 0 to 100 THz. Fig. 2 shows these two
quantities, normalized to $\rho_{sq}^2/\rho_s$, as a function of
$v_sq\tau$. They are equal in magnitude but opposite in sign,
which shows that \textit{all} of the spectral weight removed from
the condensate appears at low frequency and none appears at
$\omega_s$. Moreover, the decrease of condensate spectral weight
coincides exactly with the prediction of Barabash \textit{et al.}
\cite{bs} as $v_sq\tau\rightarrow0$, but vanishes for
$v_sq\tau\gg$1.

\begin{figure}[h]
     \includegraphics[width=3.0in]{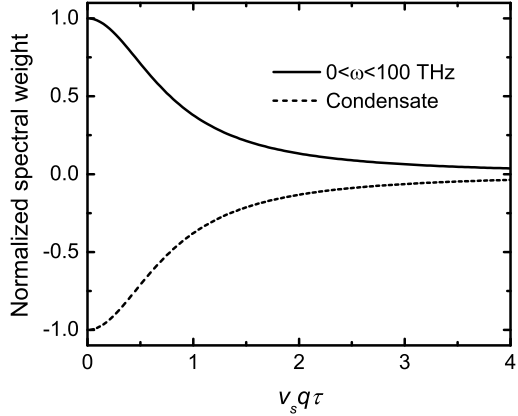}
\caption{Comparison of the spectral weight displaced from the
condensate by the quenched variation in $\rho_s$ and the spectral
weight that appears in $\Delta\sigma_1$ at low frequency.}
\label{fig:Second}
\end{figure}

The surprising results presented above are a straightforward
consequence of the anticorrelation of the density fluctuations.
There is no dissipation near $\omega_s$ or
$\omega_p\equiv[(\rho_s+\rho_n)/\epsilon_0]^{1/2}$ because the
inhomogeneous medium is \textit{dynamically} homogeneous at high
frequencies.  For $\omega\sim\omega_p$ the fractional difference
in the conductivity of the superfluid rich and superfluid poor
regions is only $\sim 1/(\omega_p\tau)^2$. The anomalous
dissipation appears instead at low frequency where the
conductivity is strongly inhomogeneous. $\sigma$ is smaller in the
regions that are superfluid poor and normal fluid rich. $E$ will
be larger in such regions, which is precisely equivalent to more
rapid phase variation in regions of low stiffness. Thus the
increase in dissipation arises ultimately from an amplification of
the electric field in regions with greater than average $\rho_n$.
Finally, the dynamical inhomogeneity disappears when $\omega\gg
v_sq$ because the normal and superfluid response again become
indistinguishable in this regime.

\begin{figure}[h]
     \includegraphics[width=3.0in]{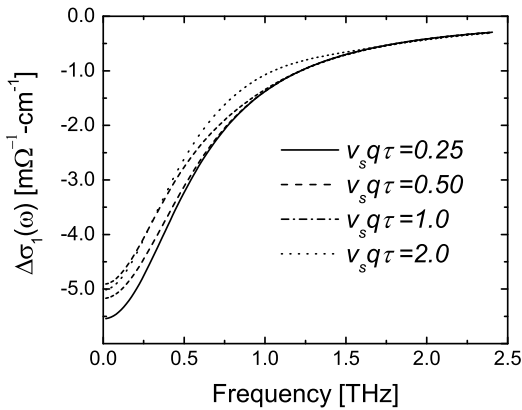}
\caption{$\Delta\sigma_1$ vs. frequency for the same parameters as
Fig. 1, except variations in $\rho_s$ and $\rho_n$ are positively
correlated.} \label{fig:Third}
\end{figure}

We are now prepared to understand the sharp difference in
$\Delta\sigma$ if the super and normal fluid density waves are in
phase.  Fig. 3 shows $Re\{\Delta\sigma(\omega)\}$ calculated with
identical parameters as in Fig. 1, except that $\rho_s=\rho_n$.
The change in conductivity is negative, and the spectra are nearly
independent of $q$ in contrast to the strongly $q$ dependent and
positive change generated when $\rho_s=-\rho_n$. The reduction in
conductivity is exactly as expected from the previous arguments:
now regions of large $E$ coincide with normal fluid poor regions
and the dissipation is attenuated.  The spectra are nearly
independent of $q$ because the dynamical inhomogeneity does not
tend to zero when the normal and superfluid response become
indistinguishable. The spectral weight is displaced equally from
the condensate and the normal fluid, and weight
2$\rho_{sq}^2/\rho_s$ shifts to $\omega_p$.

We conclude with a qualitative comparison of the calculation and
the low frequency $\sigma(\omega)$ of the cuprates. Despite an
oversimplified representation of quenched disorder, the model
provides a natural explanation for the most surprising feature of
the data, the dramatic increase of the anomalous spectral weight
as $x$ increases from under to overdoped values. The calculation
shows that this can occur if the spatial correlation of $\rho_s$
and $\rho_n$ depends strongly on $x$. Such a change in the nature
of the correlations upon crossing optimal doping $x_{opt}$ would
indeed be expected, based on very general considerations.  For a
sample with $x<x_{opt}$, regions of lower than average $\rho_s$
will be strongly underdoped patches. These patches will be
insulators, or at least regions with small $\rho_n$. On the other
hand, the regions of low $\rho_s$ in a sample with $x>x_{opt}$
will be strongly overdoped, \textit{i.e}. regions of large
$\rho_n$.  Thus the correlation varies systematically from
positive to negative with increasing $x$, leading to a dramatic
increase in the low frequency spectral weight.

We acknowledge valuable discussions with D.-H. Lee, J.C. Davis, S.
Kivelson, A.J. Millis, and P.A. Lee.  This work was supported by
NSF-9870258 and DOE-DE-AC03-76SF00098.

\end{document}